\documentstyle[epsfig,referee]{mn}

\title[]{Constraining the Cosmological Parameters and Transition
Redshift with Gamma-Ray Bursts and Supernovae}

\author[]{F. Y. Wang and Z. G. Dai
\thanks{dzg@nju.edu.cn(ZGD)}\\
{\sl Department of Astronomy, Nanjing University, Nanjing 210093,
P. R. China}}

\begin{document}

\maketitle
\begin{abstract}
A new method of measuring cosmology with gamma-ray bursts(GRBs) has
been proposed by Liang and Zhang recently. In this method, only
observable quantities including the rest frame peak energy of the
$\nu F_{\nu}$ spectrum $(E_{p}^{'})$, the isotropic energy of GRB
$(E_{\gamma,iso})$, and the break time of the optical afterglow
light curves in the rest frame $(t_{b}^{'})$ are used. By
considering this method we constrain the cosmological parameters and
the redshift at which the universe expanded from the deceleration to
acceleration phase. We add five recently-detected GRBs to the sample
and derive $E_{\gamma, {\rm{iso}}}/10^{52} {\rm
ergs}=(0.93\pm0.25)\times (E^{'}_{\rm {p}}/{\rm 100 \
keV})^{1.91\pm0.32}\times(t^{'}_{\rm{b}}/1{\rm day})^{-0.93\pm0.38}$
for a flat universe with $\Omega_M=0.28$ and $H_0=71.0$ km s$^{-1}$
Mpc$^{-1}$. This relation is independent of the medium density
around bursts and the efficiency of conversion of the explosion
energy to gamma-ray energy. We regard the $E_{\gamma,
{\rm{iso}}}(E_{\rm {p}}^{'}, t_{\rm{b}}^{'}$) relationship as a
standard candle and find $0.05<\Omega_{\rm{M}}<0.48$ and
$\Omega_{\Lambda}<1.15$ (at the $1\sigma$ confidence level). In a
flat universe with the cosmological constant we obtain
$0.25<\Omega_{\rm{M}}<0.46$ and $0.54<\Omega_{\Lambda}<0.78$ at the
$1\sigma$ confidence level. The transition redshift is
$z_{T}=0.69_{-0.12}^{+0.11}$. Combining 20 GRBs with 157 type Ia
supernovae, we find $\Omega_{M}=0.29\pm0.03$ for a flat universe and
the transition redshift is $z_{T}=0.61_{-0.05}^{+0.06}$, which is
slightly larger than the value found by considering SNe Ia alone. In
particular, We also discuss several dark-energy models in which the
equation of state $w(z)$ is parameterized, and investigate
constraints on the cosmological parameters in detail.

\end{abstract}

\begin{keywords}
cosmology: observations - distance scale - gamma-rays: bursts -
supernovae: general
\end{keywords}

\section{Introduction}
The property of dark energy and the physical cause of acceleration of the
present universe are two of the most difficult problems in modern cosmology. In
past several years, many authors used distant type Ia supernovae (SNe Ia)
(Riess et al. 1998; Perlmutter et al. 1999; Riess et al. 2004), cosmic
microwave background (CMB) fluctuations (Bennett et al. 2003; Spergel et al.
2003), and large-scale structure (LSS) (Tegmark et al. 2004) to explore
cosmology. Very recently, there have been extensive discussions on using
gamma-ray bursts (GRBs) to constrain cosmological constraints (Dai et al. 2004;
Ghirlanda et al. 2004; Xu et al. 2005; Firmani et al. 2005; Friedman \& Bloom
2005; Mortsell \& Sollerman 2005; Di Girolamo et al. 2005; Liang \& Zhang 2005;
Lamb et al. 2005).

SNe Ia have been considered as astronomical standard candles and
used to measure the geometry and dynamics of the universe. Phillips
(1993) found the intrinsic relation in SNe Ia:
$L_{p}=a\times\bigtriangleup{m}_{15}^{b}$, where $L_{p}$ is the peak
luminosity and $\bigtriangleup{m}_{15}$ is the decline rate in the
optical band at day 15 after the peak. This relation and other
similar relations can be used to explore cosmology. Riess et al
(1998) considered 16 high-redshift supernovae and 34 nearby
supernovae and found that our present universe has been
accelerating. Perlmutter et al (1999) used 42 SNe Ia and drew the
same conclusion. Riess et al (2004) selected 157 well-measured SNe
Ia, which is called the``gold" sample. Assuming a flat universe,
they concluded that: (1) Using the strong prior of
$\Omega_{M}=0.27\pm0.04$, fitting a static dark energy equation of
state yields $-1.46<w<-0.78$ (95\%C.L.). (2) Assuming a possible
redshift dependence of $w(z)$ (using $w(z)=w_{0}+w_{1}z$), the data
with the strong prior indicate that the region $w_{1}< 0$ and
especially the quadrant ($w_{0}>-1$ and $w_{1}<0$) are the least
favored. (3) Expand $q(z)$ into two terms: $q(z)=q_{0}+zdq/dz$. If
the transition redshift is defined through $q(z_{T})=0$, they found
$z_{T}=0.46\pm0.13$. The cosmological use of SNe Ia has the
following advantages: the SN Ia sample is very large and includes
low-$z$ sources, so the parameters $a$ and $b$ can be calibrated by
using low-$z$ SNe Ia. The Phillips relation and other similar
relations are intrinsic and cosmology-independent so that they can
be used to explore cosmology. But they also have disadvantages: the
interstellar medium extinction may exist when optical photons
propagate towards us. In addition, the maximum redshift of SNe Ia is
only about 1.7 and thus the earlier universe may not be
well-studied. Higher-redshift SNe Ia are necessary to eliminate
parameter degeneracies in studying the evolution of dark energy
(Weller \& Albrencht 2002; Linder \& Huterer 2003).

GRBs are the most intense electromagnetic explosions in the universe
after the big bang. They have been well understood since the
discovery of afterglows in 1997 (for review articles see Piran 1999,
2004; van Paradijs et al. 2000; M\'esz\'aros 2002; Zhang \&
M\'esz\'aros 2004). It has been widely believed that they should be
detectable out to very high redshifts (Lamb \& Reichart 2000; Ciardi
\& Loeb 2000; Bromm \& Loeb 2002; Gou et al. 2004). Schaefer (2003)
derived the luminosity distances of 9 GRBs with known redshifts by
using two quantities (the spectral lag and the variability). He
obtained the first GRB Hubble diagram with the mass density
$\Omega_M<0.35$ (at the $1\sigma$ confidence level). Ghirlanda et
al. (2004a) found the relation between isotropic-equivalent energy
$E_{\gamma,iso}$ and the local-observer peak energy $E_{p}^{'}$
(i.e., the Ghirlanda relation). Unfortunately, because of the
absence of low-$z$ GRBs, the Ghirlanda relation has been obtained
only from moderate-$z$ GRBs. So this relation is
cosmology-dependent. Dai, Liang \& Xu (2004) used for the first time
the Ghirlanda relation with 12 bursts and found the mass density
$\Omega_M=0.35\pm0.15$ (at the $1\sigma$ confidence level) for a
flat universe with the cosmological constant and the $w$ parameter
of the static dark energy model $-1.27<w<-0.50$ ($1\sigma$).
Combining 14 GRBs with SNe Ia, Ghirlanda et al. (2004b) obtained
$\Omega_{\rm M}=0.37\pm0.10$ and $\Omega_{\Lambda}=0.87\pm 0.23$.
Assuming a flat universe, the cosmological parameters were
constrained: $\Omega_{\rm M}=0.29\pm0.04$ and
$\Omega_{\Lambda}=0.71\pm 0.05$. Firmani et al (2005) used the
Bayesian method to solve the circularity problem. For a flat
universe they found $\Omega_{M}=0.28\pm0.03$ and $z_{T}=0.73\pm0.09$
for the combined GRB+SN Ia sample. In the dark energy model of
$w_{z}=w_{0}$, they found $\Omega_{M}=0.44$ and $w_{0}=-1.68$ with
$z_{T}=0.40$ for the combined GRB+SN Ia sample. In the dark energy
model of $w_{z}=w_{0}+w_{1}z$, they found the best values for GRB+SN
Ia sample were $w_{0}=-1.19$ and $w_{1}=0.98$ with $z_{T}=0.55$. Xu,
Dai \& Liang (2005) obtained
$\Omega_{M}=0.15^{+0.45}_{-0.13}(1\sigma)$ using 17 GRBs. Friedmann
\& Bloom (2005) discussed several possible sources of systematic
errors in using GRBs as standard candles. Liang \& Zhang (2005a)
presented a multi-variable regression analysis to three observable
quantities for a sample of 15 GRBs without assumption of any
theoretical models. They obtained a relation among the isotropic
gamma-ray energy ($E_{\gamma,{\rm{iso}}}$), the peak energy of the
$\nu F_{\nu}$ spectrum in the rest frame ($E_{\rm {p}}^{'}$), and
the rest frame break time of the optical afterglow light curves
($t_{\rm{b}}^{'}$). Using this relation, they found the $1\sigma$
constraints are $0.13<\Omega_{\rm{M}}<0.49$ and
$0.50<\Omega_{\Lambda}<0.85$ for a flat universe. They also obtained
the transition redshift $0.78^{+0.32}_{-0.23}$ ($1\sigma$).
Ghirlanda et al (2005) used their relation in a homogeneous density
profile and a wind density profile to explore cosmology. Liang \&
Zhang (2006) proposed an approach to calibrate the GRB luminosity
indicators without introduction of a low-redshift GRB sample. The
cosmological use of GRBs has advantages: First, gamma-ray photons
suffer from no extinction when they propagate towards us. Second,
GRBs are likely to occur at high redshifts. We can thus study the
early universe. Recently GRB 050904 whose redshift is 6.29 was
detected (Kawai et al. 2005; Haislip et al. 2005). But the low-$z$
GRB sample is so small that the intrinsic relation cannot be
obtained. A cosmology-dependent relation is now used to constrain
the cosmological parameters and transition redshift.

In this paper we investigate cosmological constraints and the
transition redshift following the method of Liang \& Zhang (2005).
Our GRB sample contains 20 bursts. Combining this sample with 157
SNe Ia we constrain the cosmological parameters and the transition
redshift in several dark energy models in detail. The structure of
this paper is arranged as follows: In section 2, we list our sample
and results from the regression analysis. In section 3, we explore
the constraints on cosmological parameters and transition redshift
using the GRB and SN Ia sample in different dark-energy models. In
section 4, we present conclusions and a brief discussion.

\section{The Method}

\subsection{Sample Selection}
Our sample includes 20 GRBs. We add GRBs 970828, 990705, 041006, 050408, and
050525a to the sample of Liang \& Zhang (2005). The redshift $z$, spectral peak
energy $E_{p}$ and optical break time $t_{b}$ of these bursts have been well
measured. The uncertainties of $E_{p}$, $S_{\gamma}$ and $k$-correction of some
bursts have not been reported. They are taken to be 20\%,10\% and 5\% of the
values. Our GRB sample is listed in Table 1.

\subsection{Cosmology with the Cosmological Constant}
The isotropic-equivalent gamma-ray energy of a GRB is calculated
by
\begin{equation}
E_{\gamma,{\rm{iso}}}=\frac{4\pi D^2_L(z)S_{\gamma}k}{1+z},
\end{equation}
where $D_{L}(z)$ is the luminosity distance at redshift $z$, and $k$ is the
factor that corrects the observed fluence to the standard rest-frame bandpass
(1-$10^{4}$ keV; Bloom et al. 2001). The expression of $D_{L}(z)$ is different
in different dark-energy models. In a Friedmann-Robertson-Walker (FRW)
cosmology with mass density $\Omega_M$ and vacuum energy density
$\Omega_\Lambda$, the luminosity distance in equation (1) is
\begin{eqnarray}
D_{L}(z) &=& c(1+z)H_0^{-1}|\Omega_k|^{-1/2}{\rm
sinn}\{|\Omega_k|^{1/2}\nonumber \\ & & \times
\int_0^zdz[(1+z)^2(1+\Omega_Mz)-z(2+z)\Omega_\Lambda]^{-1/2}\},
\end{eqnarray}
where $c$ is the speed of light and $H_0\equiv 100h\,\,{\rm
km}\,{\rm s}^{-1}\,{\rm Mpc}^{-1}$ is the present Hubble constant
(Carroll, Press \& Turner 1992). In equation (2),
$\Omega_k=1-\Omega_M-\Omega_\Lambda$, and ``sinn" is sinh for
$\Omega_k>0$ and sin for $\Omega_k<0$. For $\Omega_k=0$, equation
(2) turns out to be $c(1+z)H_0^{-1}$ times the integral. In this
model, the transition redshift satisfies
\begin{equation}
z_{T}=\left(\frac{2\Omega_{\Lambda}}{\Omega_{M}}\right)^{1/3}-1.
\end{equation}

\subsection{One-parameter dark-energy model}
We consider an equation of state for dark energy
\begin{equation}
w_{z}=w_{0}.
\end{equation}
In this dark energy model, the luminosity distance for a flat
universe is (Riess et al. 2004)
\begin{equation}
D_{L}=cH_{0}^{-1}(1+z)\int_{0}^{z}dz[(1+z)^{3}\Omega_{M}+(1-\Omega_{M})(1+z)^{3(1+w_{0})}]^{-1/2}.
\end{equation}
The transition redshift satisfies
\begin{equation}
\Omega_{M}+(1-\Omega_{M})(1+3w_{0})\times(1+z)^{3w_{0}}=0.
\end{equation}

\subsection{Two-parameter dark-energy model}
A more interesting approach to explore dark energy is to use a
time-dependent dark energy model. The simplest parameterization
including two parameters is (Maor et al. 2001; Weller \& Albrecht
2001, 2002; Riess et al 2004)
\begin{equation}
w_{z}=w_{0}+w_{1}z.
\end{equation}
This parameterization provides the minimum possible resolving power
to distinguish between the cosmological constant and a rolling
scalar field from  the time variation. The value of $w_{1}$ could
provide an estimate of the scale length of a dark energy potential
(Riess et al 2004). In this dark energy model the luminosity
distance is (Linder 2003)
\begin{equation}
D_{L}=cH_{0}^{-1}(1+z)\int_{0}^{z}dz[(1+z)^{3}\Omega_{M}+(1-\Omega_{M})(1+z)^{3(1+w_{0}-w_{1})}e^{3w_{1}z}]^{-1/2}.
\end{equation}
The transition redshift is given by
\begin{equation}
\Omega_{M}+(1-\Omega_{M})(1+3w_{0}+3w_{1}z)\times(1+z)^{w_{0}-w_{1}}e^{3w_{1}z}=0.
\end{equation}

The above model is incompatible with the CMB data since it diverges at high
redshifts (Chevallier \& Polarski 2001). Linder (2003) proposed an extended
parameterization which avoids this problem,
\begin{equation}
w_{z}=w_{0}+\frac{w_{1}z}{1+z}.
\end{equation}
We adopt the results only if $w_{0}+w_{1}$ is well below zero at the time of
decoupling. In this dark energy model the luminosity distance is
\begin{equation}
D_{L}=cH_{0}^{-1}(1+z)\int_{0}^{z}dz[(1+z)^{3}\Omega_{M}+(1-\Omega_{M})(1+z)^{3(1+w_{0}+w_{1})}e^{-3w_{1}z/(1+z)}]^{-1/2}.
\end{equation}
The transition redshift is given by
\begin{equation}
\Omega_{M}+(1-\Omega_{M})\left(1+3w_{0}+\frac{3w_{1}z}{1+z}\right)\times(1+z)^{w_{0}+w_{1}}e^{-3w_{1}z/(1+z)}=0.
\end{equation}

By fitting the SN Ia data using the $w_{z}=w_{0}+\frac{w_{1}z}{1+z}$
model, $w_{0}+w_{1}>0$ was found and thus at high redshifts this
model is not good. In order to solve this problem, Jassal, Bagla and
Padmanabhan modified this parameterization as
\begin{equation}
w_{z}=w_{0}+\frac{w_{1}z}{(1+z)^2}.
\end{equation}
This equation can model a dark energy component which has the same
value at lower and higher redshifts, with rapid variation at low
$z$. Observations are not very sensitive to variations in $w_{z}$
for $z\gg1$. However, it does allow us to probe rapid variations
at small redshifts (Jassal, Bagla \& Padmanabhan 2004). The
luminosity distance in this dark energy model is
\begin{equation}
D_{L}=cH_{0}^{-1}(1+z)\int_{0}^{z}dz[(1+z)^{3}\Omega_{M}+(1-\Omega_{M})(1+z)^{3(1+w_{0})}e^{3w_{1}z^{2}/2(1+z)^{2}}]^{-1/2}.
\end{equation}
The transition redshift is given by
\begin{equation}
\Omega_{M}+(1-\Omega_{M})\left(1+3w_{0}+\frac{3w_{1}z}{(1+z)^{2}}\right)\times(1+z)^{3w_{0}}e^{3w_{1}z^{2}/2(1+z)^{2}}=0.
\end{equation}

\subsection{Regression Analysis}
We perform a three-variable regression analysis to find an empirical
relation among $E_{\gamma, {\rm{iso}}}$, $E_{\rm {p}}^{'}$, and
$t_{\rm{b}}^{'}$. The model that we use is
\begin{equation}\label{MVR}
\hat{E}_{{\rm{iso}}}=10^{\kappa_{0}} E_{p}^{'\kappa_{1}} t_{b}^{'
\kappa_{2}},
\end{equation}
where $E_{p}^{'}=E_{\rm {p}}(1+z)$ and $t_{b}^{'}=t_{\rm{b}}/(1+z)$. For a flat
universe with $\Omega_{M}=0.28$, using the sample of 20 GRBs, we find a
relation among $E_{\gamma, {\rm{iso}}}$, $E_{\rm {p}}^{'}$, and
$t_{\rm{b}}^{'}$,
\begin{eqnarray}\label{MR}
\hat{E}_{\gamma,{\rm{iso,52}}}&=&(0.93\pm0.25)\times
\left(\frac{E^{'}_{\rm p}}{100 ~{\rm
keV}}\right)^{1.91\pm0.32}\\
&&\times  \left(\frac{t'_{\rm b}}{1~{\rm
day}}\right)^{-0.93\pm0.38},
\end{eqnarray}
where $\hat{E}_{\gamma,{\rm{iso,52}}}=\hat{E}_{\gamma, {\rm{iso}}}/10^{52} {\rm
ergs}$ (see Figure 1). This relation depends on the cosmology that we choose.
The dispersion of this relation is so small that it can be used to study
cosmology (Liang \& Zhang 2005). We don't assume any theoretical models when
deriving this relation. With $D_{L}$ in units of mega parsecs, the distance
modulus is
\begin{equation}
\hat\mu=5\log(D_{L})+25.
\end{equation}
Thus,
\begin{eqnarray}\label{mu}
\hat{\mu}&=&2.5[\kappa_0+\kappa_1\log E_{p}^{'}+\kappa_2\log t_{b}^{'}\\
&& -\log ({4\pi S_{\gamma}k})+\log ({1+z})]-97.45.
\end{eqnarray}
Because of the cosmology-dependent relation, $\hat\mu$ is
cosmology-dependent. We need low-$z$ GRBs to calibrate a
cosmology-independent relation. But the current low-$z$ GRBs sample
is small.

We use the following method to solve this problem (also see Liang
\& Zhang 2005):

(1) Given a particular set of cosmological parameters
($\bar\Omega$), we calculate the correlation
$\hat{E}_{\gamma,\rm{iso}}(\bar{\Omega}; E_{p}^{'},t_{b}^{'})$. We
evaluate the probability ($w(\bar{\Omega})$) of using this
relation as a cosmology-independent luminosity indicator via
$\chi^2$ statistics, i.e.,
\begin{equation}\label{chir}
\chi^2_{w}(\bar{\Omega})=\sum_{i}^{N}\frac{[\log
\hat{E}^{i}_{\gamma, {\rm iso}}(\bar{\Omega})-\log E^{i}_{\gamma,
{\rm iso}}(\bar{\Omega})]^2}{\sigma_{\log \hat{E}^{i}_{\gamma,
{\rm iso}}}^2(\bar{\Omega})}.
\end{equation}
The probability is
\begin{equation}\label{weight}
w(\bar{\Omega})\propto e^{-\chi^2_{w}(\bar{\Omega})/2}.
\end{equation}

(2) Regard the relation derived in step (1) as a
cosmology-independent luminosity indicator without considering its
systematic error, and calculate distance modulus
$\hat{\mu}(\bar{\Omega})$ and its error $\sigma_{\hat{\mu}}$,
\begin{eqnarray}\label{err2}
\sigma_{\hat{\mu}_{i}}&=&\frac{2.5}{\ln 10}[(\kappa_1
\frac{\sigma_{E_{\rm p,i}^{'}}}{ E_{\rm
p,i}^{'}})^2+(\kappa_2\frac{\sigma_{t_{\rm b,i}^{'}}}{t_{\rm
b,i}^{'}})^2
+(\frac{\sigma_{S_{\gamma,i}}}{S_{\gamma,i}})^2\\
&&+(\frac{\sigma_{k_i}}{k_i})^2+(\frac{\sigma_{z_i}}{1+z_{i}})^2]^{1/2}.
\end{eqnarray}

(3) Derive the theoretical distance modulus $\mu (\Omega)$ in a
set of cosmological parameters, and the $\chi^2$ of $\mu (\Omega)$
against $\hat{\mu}(\Omega)$,
\begin{equation}\label{chis}
\chi^2(\bar{\Omega}|\Omega)=\sum_{i}^{N}
\frac{[\hat{\mu}_i(\bar{\Omega})-\mu_i(\Omega)]^2}{\sigma_{
\hat{\mu}_{i}}^2(\bar{\Omega})}.
\end{equation}

(4) Calculate the probability that the cosmology parameter set
$\Omega$ is the right one according to the luminosity indicator
derived from the cosmological parameter set $\bar{\Omega}$,
\begin{equation}\label{pp}
p(\bar{\Omega}|\Omega)\propto e^{ -\chi^2(\bar{\Omega}|\Omega)/2}.
\end{equation}

(5) Integrate $\bar{\Omega}$ over the full cosmological parameter
space to get the final normalized probability that the cosmological
$\Omega$ is the right one,
\begin{equation}\label{PW}
p(\Omega)=\frac{\int_{\bar{\Omega}}w(\bar{\Omega})
p(\bar{\Omega}|\Omega)d\bar{\Omega}}{\int_{\bar{\Omega}}
w(\bar{\Omega})d\bar{\Omega}}.
\end{equation}

In our calculation, the integration in eq.(\ref{PW}) is computed
through summing over a wide range of the cosmology parameter space
to make the sum converge, i.e.,
\begin{equation}\label{PWsum}
p(\Omega)=\frac{\sum_{\bar{\Omega}_i}w(\bar{\Omega}_i)
p(\bar{\Omega}_i|\Omega)}{\sum_{\bar{\Omega}_i}w(\bar{\Omega}_i)}.
\end{equation}

\section{Results}
\subsection{Cosmology with the Cosmological Constant}
We obtain Figure 2 in a Friedmann-Robertson-Walker (FRW) cosmology
with mass density $\Omega_M$ and vacuum energy density
$\Omega_\Lambda$ using the GRB sample. This figure shows that
$0.05<\Omega_{\rm{M}}<0.48$ at the $1\sigma$ confidence level, but
$\Omega_\Lambda$ is poorly constrained, $\Omega_{\Lambda}<1.15$ .
For a flat universe, we obtain $0.25<\Omega_{\rm{M}}<0.46$ and
$0.54<\Omega_{\Lambda}<0.78$ at the $1\sigma$ confidence level. The
best values of $(\Omega_{\rm{M}},\Omega_{\Lambda})$ are $(0.29,
0.62)$. The transition redshift that the universe changed from the
deceleration to acceleration phase is $z_{T}=0.69_{-0.12}^{+0.11}$
at the $1\sigma$ confidence level.

We plot Figure 3 combining the GRB sample with the SN Ia sample.
From this figure we find $\Omega_{M}=0.29\pm0.03$ at the $1\sigma$
confidence level for a flat universe. The gray contours are derived
from SNe Ia alone. The dashed contours are derived from the GRB and
SNe Ia sample. We can see that the contours move down and become
tighter as compared with the case of a combination of GRBs and SNe
Ia. The result is more convincing than the one from SNe Ia alone.
The transition redshift is $z_{T}=0.61_{-0.05}^{+0.06}$, which is
slightly larger than the value found by considering SNe Ia alone.
Figure 4 shows the confidence level derived from SN Ia sample and
eight $z>1.5$ GRBs. The results are similar to those shown in Figure
3. We can conclude that an improvement of the confidence is due to
the contribution of GRBs at moderate redshifts.

\subsection{One-parameter dark-energy model}
Figure 5 shows the constraints on $w_{0}$ and $\Omega_{M}$ in this dark energy
model. The best values for the SN Ia and GRB sample are
$\Omega_{M}=0.48_{-0.09}^{+0.07}$ and $w_{0}=-1.90^{+0.58}_{-2.85}$. The
transition redshift is $z_{T}=0.50_{-0.10}^{+0.09}$. Riess et al (2004) gave
$w_{0}=-1.02_{-0.19}^{+0.13}$ using SNe Ia sample with a prior of
$\Omega_{M}=0.27\pm0.04$. Firmani et al (2005) obtained $\Omega_{M}=0.44$ and
$w_{0}=-1.68$ from their SNe Ia and GRB sample using the Bayesian method.

\subsection{Two-parameter dark-energy model}
Figure 6 exhibits the constraints on the dark energy parameters
($w_{0}$ and $w_{1}$) and transition redshift in the model of
$w_{z}=w_{0}+w_{1}z$. For the SN Ia and GRB sample the best values
are $w_{0}=-1.20_{-0.24}^{+0.18}$ and $w_{1}=0.85_{-0.25}^{+0.30}$
with $z_{T}=0.44_{-0.08}^{+0.09}$ at the $1\sigma$ confidence level.
Firmani et al (2005) obtained $w_{0}=-1.19$ and $w_{1}=0.98$ with
$z_{T}=0.55$ using their SN Ia and GRB sample. From this figure, we
see that the contours plotted from the SN Ia and GRB sample move
down and become tighter.

Figure 7 presents the constraints on the dark energy parameters ($w_{0}$ and
$w_{1}$) and transition redshift in the model of $w_{z}=w_{0}+w_{1}z/(1+z)$.
The best values for the SN Ia and GRB sample are $w_{0}=-1.12_{-0.30}^{+0.32}$
and $w_{1}=1.90_{-1.75}^{+2.10}$ (at the $1\sigma$ confidence level)with
$z_{T}=0.49_{-0.09}^{+0.07}$. From this figure we also can see that the
contours plotted from the SN Ia and GRB sample move down and become tighter,
being similar to Figure 4.

Figure 8 shows the constraints on the dark energy parameters
($w_{0}$ and $w_{1}$) and transition redshift in the model of
$w_{z}=w_{0}+w_{1}z/(1+z)^2$. The best values for the SN Ia and
GRB sample are $w_{0}=-1.00_{-0.25}^{+0.22}$ and
$w_{1}=0.54_{-1.20}^{+0.85}$ with $z_{T}=0.37_{-0.07}^{+0.10}$ at
the $1\sigma$ confidence level.

\section{CONCLUSIONS and DISCUSSION}
Following Liang \& Zhang (2005), we have derived an empirical
relation among the rest frame peak energy of the $\nu F_{\nu}$
spectrum $(E_{p}^{'})$, the isotropic energy of GRB
$(E_{\gamma,iso})$, and the break time of the optical afterglow
light curves in the rest frame $(t_{b}^{'})$ without any theoretical
assumptions. The relation is $E_{\gamma, {\rm{iso}}}/10^{52} {\rm
ergs}=(0.93\pm0.25)\times (E^{'}_{\rm {p}}/{\rm 100 \
keV})^{1.91\pm0.32}\times(t^{'}_{\rm{b}}/1{\rm day})^{-0.93\pm0.38}$
in a flat universe with $\Omega_M=0.28$ and $H_0=71.0$ km s$^{-1}$
Mpc$^{-1}$ for our sample with 20 GRBs. We find
$0.05<\Omega_{\rm{M}}<0.48$ and $\Omega_{\Lambda}<1.15$ ($1\sigma$).
For a flat universe with the cosmological constant we obtain
$0.25<\Omega_{\rm{M}}<0.46$ and $0.54<\Omega_{\Lambda}<0.78$ at the
$1\sigma$ confidence level. The transition redshift is
$z_{T}=0.69_{-0.12}^{+0.11}$. Because of the small number of useful
GRBs, we combine GRBs with SNe Ia. Using this joint sample, we find
$\Omega_{M}=0.29\pm0.03$ for a flat universe. The transition
redshift is $z_{T}=0.61_{-0.05}^{+0.06}$, which is slightly larger
than the value found by considering SN Ia alone. The best values for
$\Omega_{M}$ and $w_{0}$ are $\Omega_{M}=0.48$ and $w_{0}=-1.90$
with the GRB and SN Ia sample in the $w_{z}=w_{0}$ model. The
transition redshift is $z_{T}=0.50_{-0.10}^{+0.09}$. For the SN Ia
and GRB sample the best values are $w_{0}=-1.20_{-0.24}^{+0.18}$ and
$w_{1}=0.85_{-0.25}^{+0.30}$ with $z_{T}=0.44_{-0.08}^{+0.09}$ at
the $1\sigma$ confidence level in the model of $w_{z}=w_{0}+w_{1}z$.
The best values for the SN Ia and GRB sample are
$w_{0}=-1.12_{-0.30}^{+0.32}$ and $w_{1}=1.90_{-1.75}^{+2.10}$ (at
the $1\sigma$ confidence level) with $z_{T}=0.49_{-0.09}^{+0.07}$ in
the model of $w_{z}=w_{0}+w_{1}z/(1+z)$. The best values for the SN
Ia and GRB sample are $w_{0}=-1.00_{-0.25}^{+0.22}$ and
$w_{1}=0.54_{-1.20}^{+0.85}$ with $z_{T}=0.37_{-0.07}^{+0.10}$ at
the $1\sigma$ confidence level in the model of
$w_{z}=w_{0}+w_{1}z/(1+z)^2$. The contours are tighter than those
only by using the SN Ia data. From these figures a clear trend can
be seen.

GRBs appear to be natural events of studying the universe at very
high redshifts. They can bridge the gap between the nearby SNe Ia
and the cosmic microwave background. GRBs establish a new insight on
the cosmic acceleration. However the density of the circumburst
medium and the efficiency of conversion of the explosion energy to
gamma-rays should be assumed in the Ghirlanda relation. This is
naturally overcome in the Liang \& Zhang relation.

Since the empirical relation of Liang \& Zhang (2005) is
cosmology-dependent, we use a strategy through weighing this
relation in all possible cosmological models. We find that the
transition redshift varies from $0.37_{-0.07}^{+0.10}$ to
$0.69_{-0.06}^{+0.05}$. Although our constraints from the GRB sample
are weaker than those from SNe Ia, the constraints from the GRB and
SN Ia sample are more stringent than those from the SN Ia sample. A
low-$z$ GRB sample is needed to calibrate the relation in a
cosmology-independent way. MIDEX-class mission, which would obtain
$>$ 800 bursts in the redshift range $0.1\leq z \leq10$ during a
2-year mission (Lamb et al. 2005), will be dedicated to using GRBs
to constrain the properties of dark energy. This burst sample would
enable both $\Omega_{M}$ and $w_{0}$ to be determined to $\pm 0.07$
and $\pm 0.06$ (68\% C.L.), respectively, and $w_a$ to be
significantly constrained. Probing the properties of dark energy by
using GRBs is complementary (in the sense of parameter degeneracies)
to other probes, such as CMB anisotropies and X-ray clusters (Lamb
et al 2005). New constraints from GRBs detected in the future would
improve the study of cosmology.

We thank Enwei Liang and Bing Zhang for helpful discussions, and
the referee for valuable suggestions. This work was supported by
the National Natural Science Foundation of China (grants 10233010
and 10221001).

\newpage

\begin{table*}
\caption{The parameters of the GRB sample used in this paper}
\begin{tabular}{lllllllll}
\hline\hline%

GRB&$z$&$E_{\rm {p}}(\sigma_{E_{\rm
{p}}})$&$\alpha$&$\beta$&$S_{\gamma}(\sigma_S)$&Band &$t_{b}(\sigma_{t_{b}})$ &Refs.\\
(1)&(2)&(3)(keV)&(4)&(5)&(6)(erg.cm$^{-2}$)&(7)(keV)&(8)(days)&(9)\\

\hline

970828&0.9578&297.7(59.5)&-0.70&-2.07&96.0(9.6)&20-2000&2.2(0.4)&1; 5; 20; 20\\
980703&0.966&254(50.8)&-1.31&-2.40&22.6(2.3)&20-2000&3.4(0.5)&2; 3; 3; 3\\
990123&1.6&780.8(61.9)&-0.89&-2.45&300(40)&40-700&2.04(0.46)&4; 5; 5; 5\\
990510&1.62&161.5(16.1)&-1.23&-2.70&19(2)&40-700&1.6(0.2)&6; 5; 5; 5\\
990705&0.8424&188.8(15.2)&-1.05&-2.20&75.0(8.0)&40-700&1.0(0.2)&1; 19; 20; 20\\
990712&0.43&65(11)&-1.88&-2.48&6.5(0.3)&40-700&1.6(0.2)&6; 5; 5; 5\\
991216&1.02&317.3(63.4)&-1.23&-2.18&194(19)&20-2000&1.2(0.4)&7; 3; 3; 3 \\
011211&2.14&59.2(7.6)&-0.84&-2.30&5.0(0.5)&40-700&1.56(0.02)&8; 9; 9; 8\\
020124&3.2&86.9(15.0)&-0.79&-2.30&8.1(0.8)&2-400&3(0.4)&10; 11; 11; 11\\
020405&0.69&192.5(53.8)&0.00&-1.87&74.0(0.7)&15-2000&1.67(0.52)&12; 12; 12; 12\\
020813&1.25&142(13)&-0.94&-1.57&97.9(10)&2-400&0.43(0.06)&13; 11; 11; 11\\
021004&2.332&79.8(30)&-1.01&-2.30&2.6(0.6)&2-400&4.74(0.14)&14; 11; 11; 11\\
021211&1.006&46.8(5.5)&-0.86&-2.18&3.5(0.1)&2-400&1.4(0.5)&15; 11; 11; 11\\
030226&1.986&97(20)&-0.89&-2.30&5.61(0.65)&2-400&1.04(0.12)&16; 11; 11; 11\\
030328&1.52&126.3(13.5)&-1.14&-2.09&37.0(1.4)&2-400&0.8(0.1)&17; 11; 11; 11\\
030329&0.1685&67.9(2.2)&-1.26&-2.28&163(10)&2-400&0.5(0.1)& 18; 11; 11; 11\\
030429&2.6564&35(9)&-1.12&-2.30&0.85(0.14)&2-400&1.77(1)&19; 11; 11; 11\\
041006&0.716&63.4(12.7)&-1.37&-2.30&19.9(1.99)&25-100&0.16(0.04)&20; 20; 21; 21\\
050408&1.2357&19.93(4.0)&-1.979&-2.30&1.90(0.19)&30-400&0.28(0.17)&22; 22; 23; 23\\
050525a&0.606&79.0(3.5)&-0.987&-8.839&20.1(0.50)&15-350&0.28(0.12)&24; 24; 24; 24\\
\hline
\end{tabular}

\noindent {References are in order for $z$, $E_{\rm p}^{obs}$,
$[\alpha, \beta]$, $S_{\gamma}$, $t_{b}$:(1) Bloom et al. 2003;
(2) Djorgovski et al. 1998; (3) Jimenez et al. 2001; (4) Kulkarni
et al. 1999; (5) Amati et al. 2002; (6) Vreeswijk et al. 2001;
(7)Djorgovski et al. 1999; (8) Holland et al. 2002; (9) Amati
2004; (10) Hjorth et al. 2003; (11)Sakamoto et al. 2005; (12)
Price et al. 2003; (13) Barth et al. 2003; (14) M\"oller et al.
2002; (15) Vreeswijk et al. 2003; (16) Greiner et al. 2003; (17)
Martini et al.  2003; (18)Bloom et al. 2003c; (19) Weidinger et
al. 2003; (20) Butler et al. 2005. (21) Stanek et al. 2005. (22)
Sakamoto et al. 2005. (23) Godet et al. 2005. (24)Blustin et al.
2005.}
\end{table*}
\begin{figure}
  \begin{center}
  \centerline{ \hbox{ \epsfig{figure=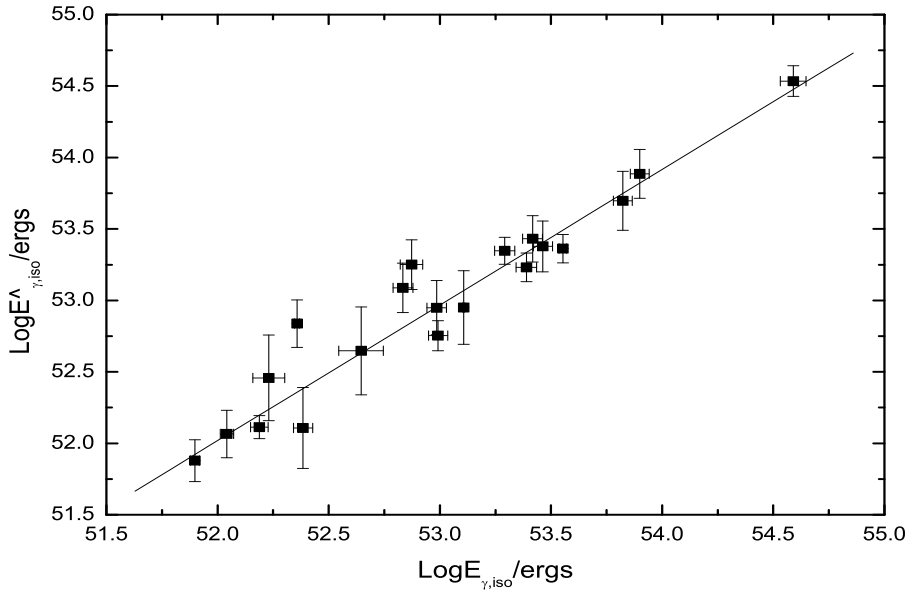,width=5.5in,height=3.8in,angle=0}}}
  \caption{log $E\hat{}_{\gamma,iso}$ calculated by the regression method versus log $E_{\gamma,iso}$
  calculated from a flat universe with $\Omega_{M}=0.28$. The line is the best
regression line.}
  \end{center}
  \end{figure}

\newpage
\begin{figure}
  \begin{center}
  \centerline{ \hbox{\epsfig{figure=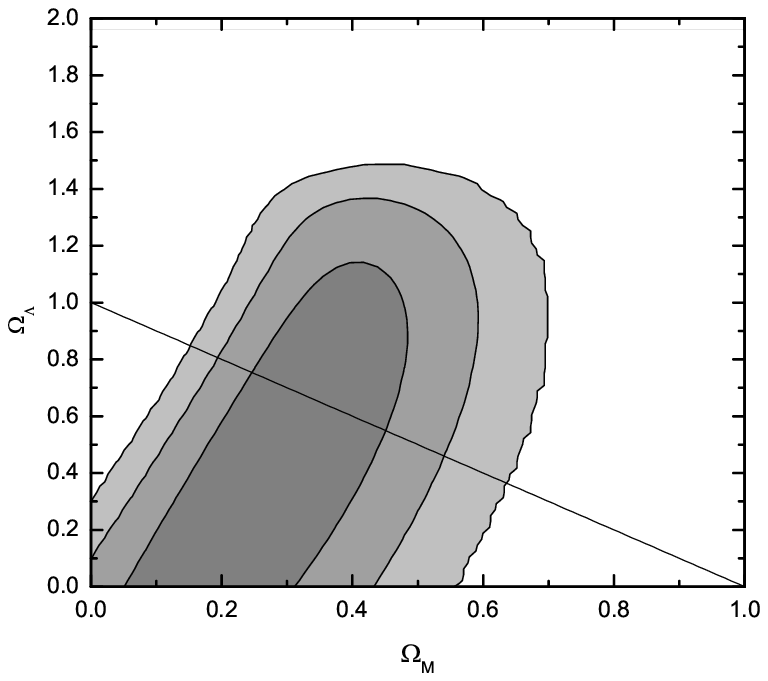,width=5.5in,height=3.8in,angle=0}}}
  \centerline{ \hbox{\epsfig{figure=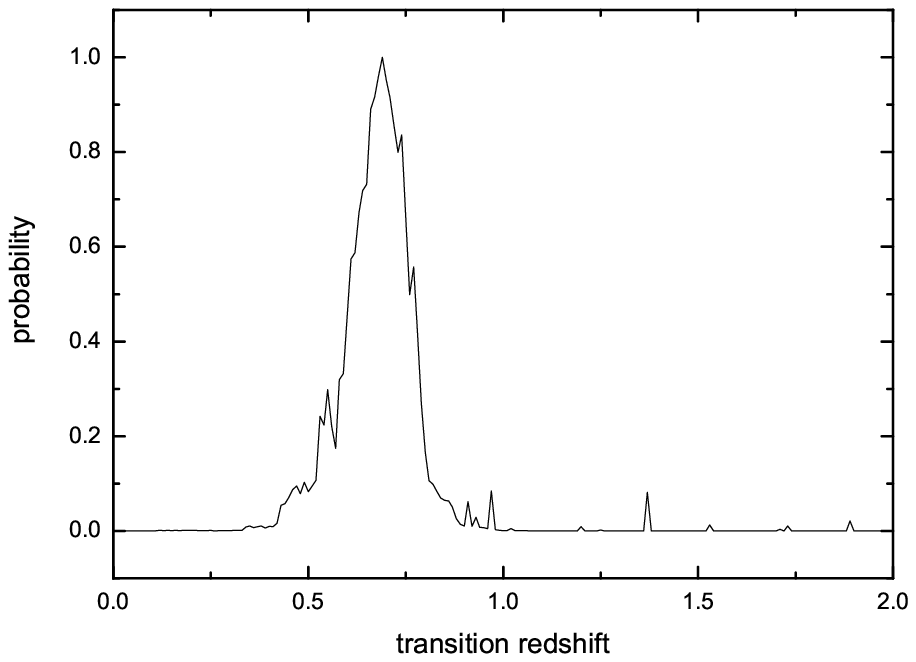,width=5.5in,height=3.8in,angle=0}}}
  \caption{The top panel shows confidence interval distributions in the $\Omega_{M}-\Omega_{\Lambda}$ plane
   from $1\sigma$ to $3\sigma$ in an FRW cosmology.
   The straight line represents the flat universe. The bottom panel shows the probability versus transition redshift.}
  \end{center}
  \end{figure}

\newpage
\begin{figure}
  \begin{center}
  \centerline{ \hbox{\epsfig{figure=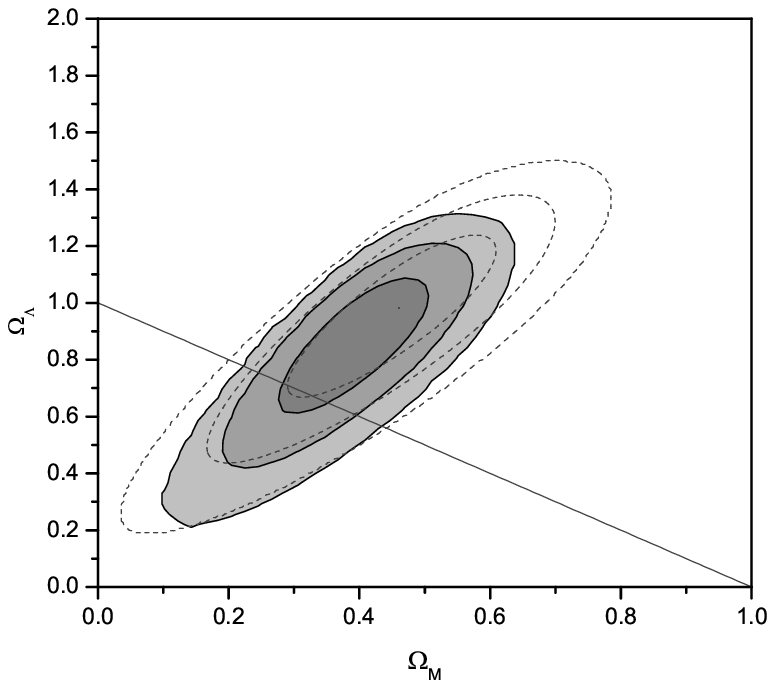,width=5.5in,height=3.8in,angle=0}}}
  \centerline{ \hbox{\epsfig{figure=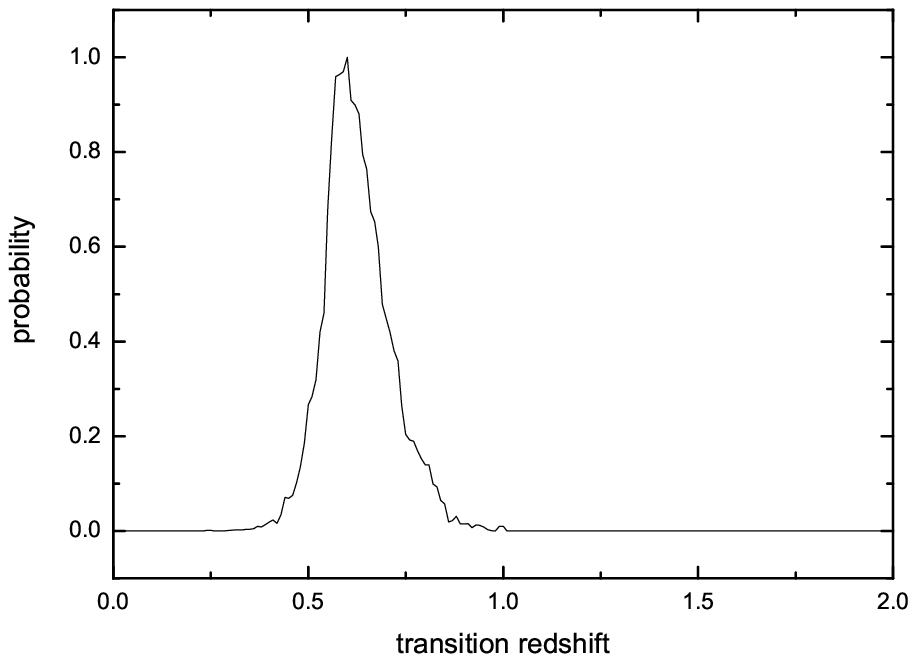,width=5.5in,height=3.8in,angle=0}}}
  \caption{Constraints on $\Omega_{M}$ and $\Omega_{\Lambda}$ from $1\sigma$ to $3\sigma$ in the top panel.
  The dashed contours are derived from 157 SNe Ia alone and the gray ones from the GRB and SN Ia sample.
  The bottom panel shows the probability versus the transition redshift derived from the GRB and SN Ia sample.}
  \end{center}
  \end{figure}

\newpage
\begin{figure}
  \begin{center}
  \centerline{ \hbox{\epsfig{figure=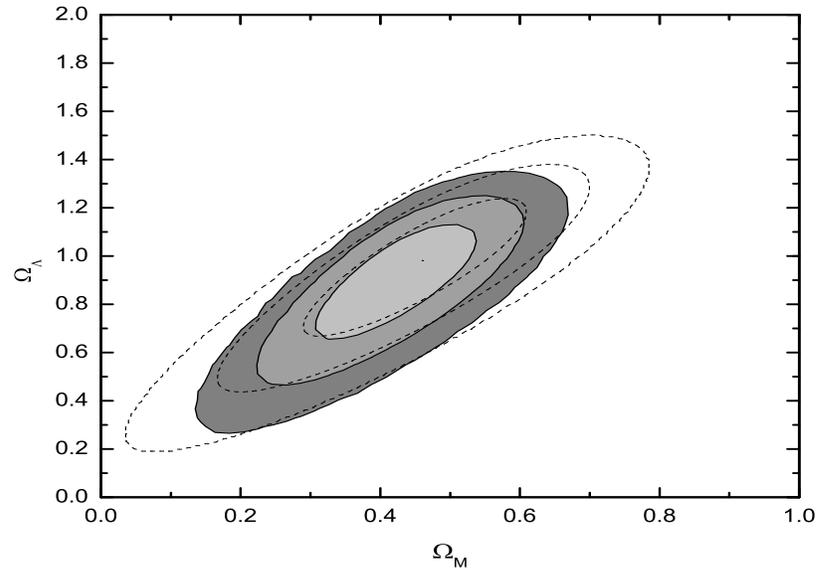,width=5.5in,height=3.8in,angle=0}}}
   \caption{Constraints on $\Omega_{M}$ and $\Omega_{\Lambda}$ from $1\sigma$ to $3\sigma$ derived from
    157 SNe Ia and 8 GRBs with redshifts greater than 1.5 (shown by gray contours).
    The dashed contours are derived from 157 SNe Ia alone.}
  \end{center}
  \end{figure}

\newpage
\begin{figure}
  \begin{center}
  \centerline{ \hbox{\epsfig{figure=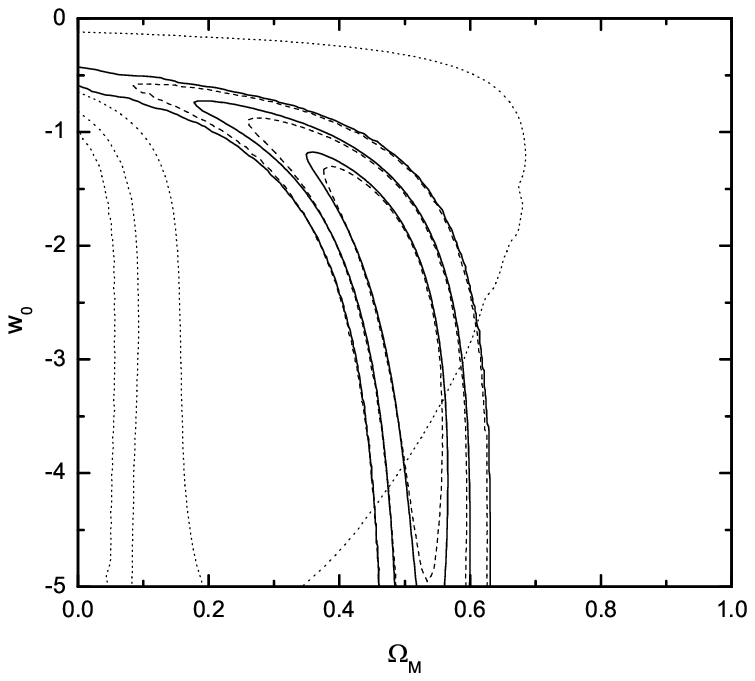,width=5.5in,height=3.8in,angle=0}}}
  \centerline{ \hbox{\epsfig{figure=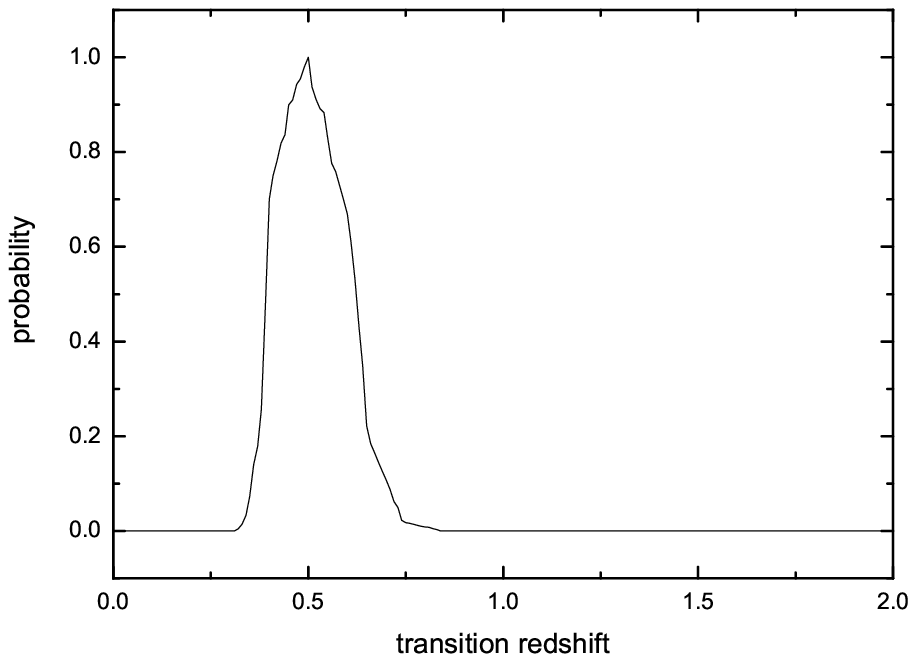,width=5.5in,height=3.8in,angle=0}}}
  \caption{ Constraints on $\Omega_{M}$ and $w_{0}$ from $1\sigma$ to $3\sigma$ with dark energy whose equation
  state is constant in the top panel.
  The solid contours are derived from SNe Ia alone and the long dashed contours
  are
derived from 20 GRBs and Gold sample. The dotted contours are derived from 20
GRBs. The bottom panel shows the probability versus transition redshift derived
from the GRB and SN Ia sample.}
  \end{center}
  \end{figure}

\newpage
\begin{figure}
  \begin{center}
  \centerline{ \hbox{ \epsfig{figure=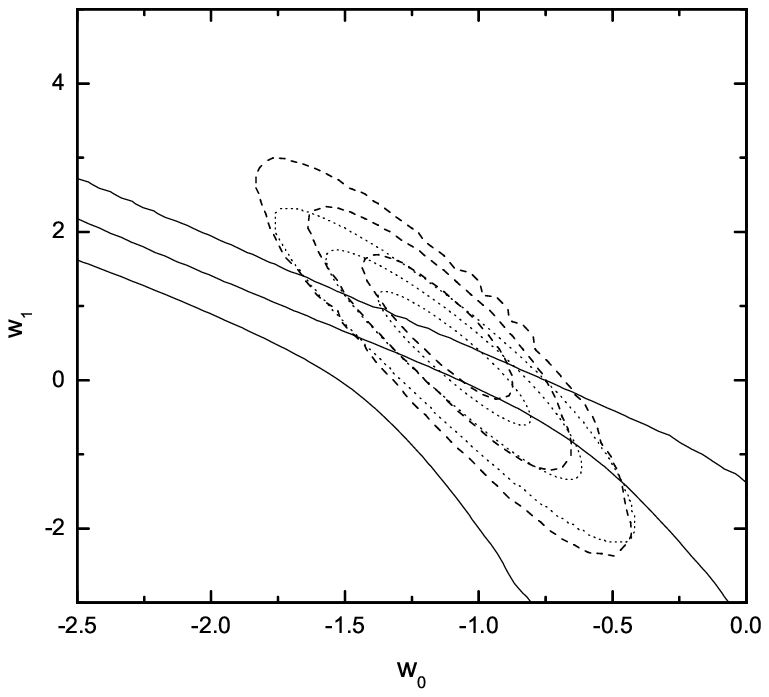,width=5.5in,height=3.8in,angle=0}}}
  \centerline{ \hbox{ \epsfig{figure=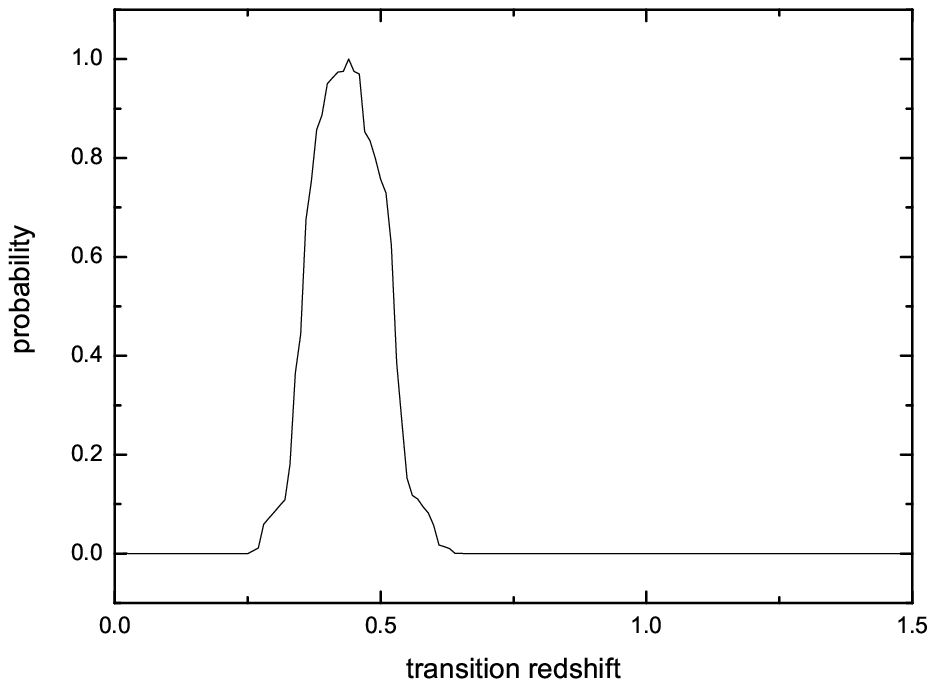,width=5.5in,height=3.8in,angle=0}}}
  \caption{Constraints on $w_{0}$ and $w_{1}$ from $1\sigma$ to $3\sigma$ with dark energy whose equation
  state is $w_{z}=w_{0}+w_{1}z$ in the top panel.
  The long dashed contours are derived from SNe Ia alone and the dotted contours
  are
derived from 20 GRBs and Gold sample. The diagonal lines are
obtained from 20 GRBs. The bottom panel shows the probability versus
transition redshift derived from the GRB and SN Ia sample.}
  \end{center}
  \end{figure}

\newpage
\begin{figure}
  \begin{center}
  \centerline{ \hbox{ \epsfig{figure=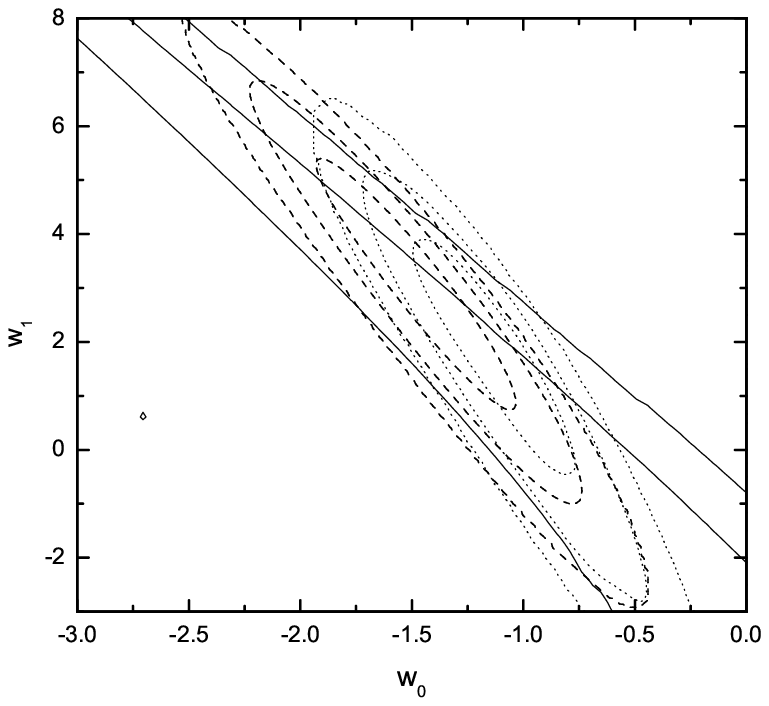,width=5.5in,height=3.8in,angle=0}}}
  \centerline{ \hbox{ \epsfig{figure=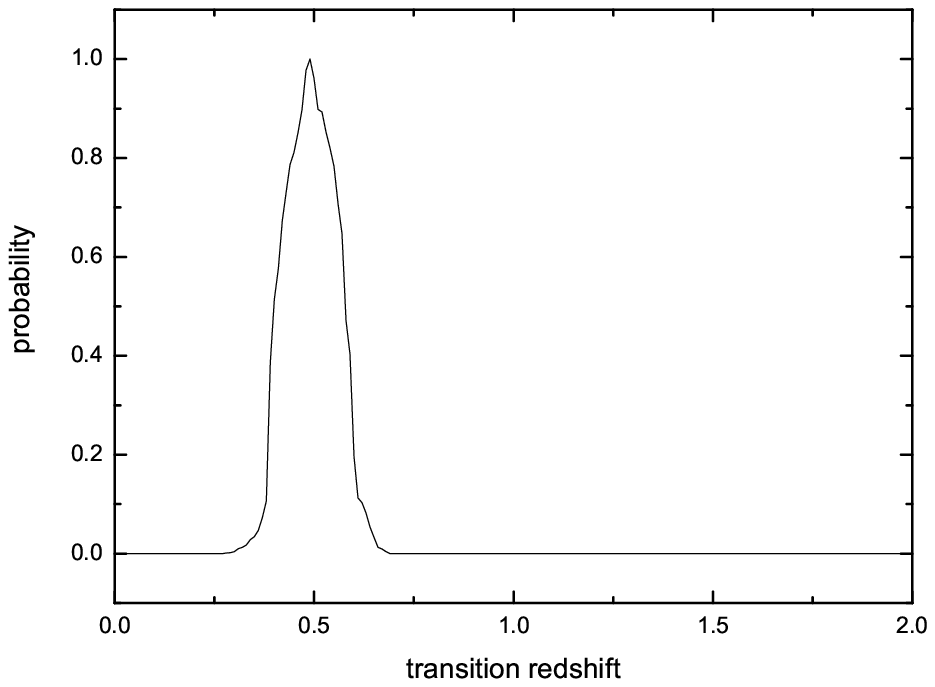,width=5.5in,height=3.8in,angle=0}}}
  \caption{Constraints on $w_{0}$ and $w_{1}$ from $1\sigma$ to $3\sigma$ with dark energy whose equation
  state is $w_{z}=w_{0}+w_{1}z/(1+z)$ in the top panel.
  The long dashed contours are derived from SNe Ia alone and the dotted contours
  are
derived from 20 GRBs and Gold sample. The diagonal lines are
obtained from 20 GRBs. The bottom panel shows the probability versus
transition redshift derived from the GRB and SN Ia sample.}
  \end{center}
  \end{figure}

\newpage
\begin{figure}
  \begin{center}
  \centerline{ \hbox{ \epsfig{figure=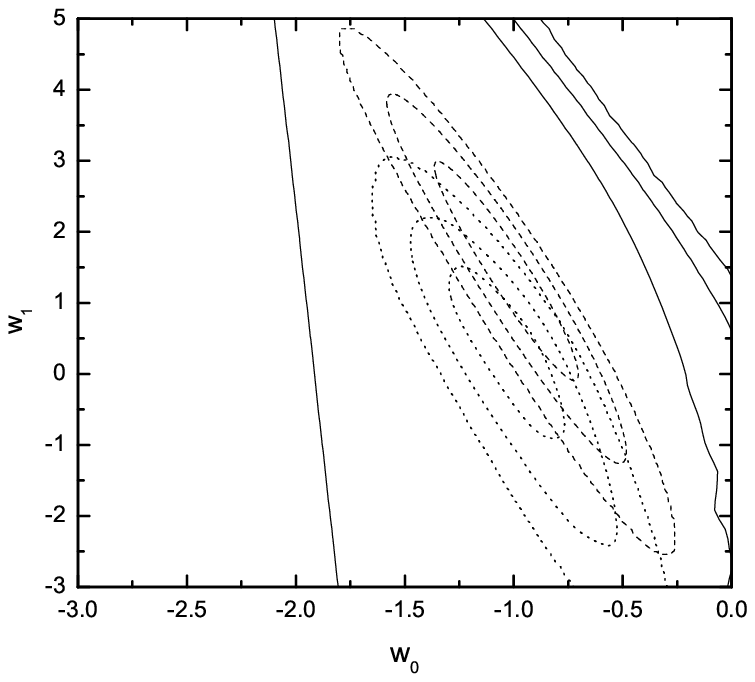,width=5.5in,height=3.8in,angle=0}}}
  \centerline{ \hbox{ \epsfig{figure=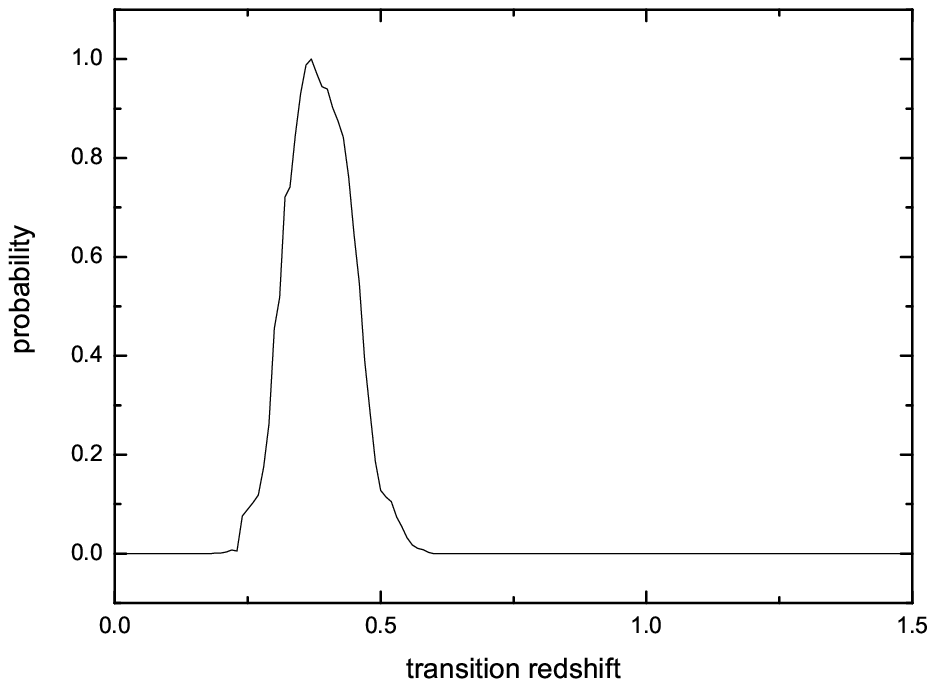,width=5.5in,height=3.8in,angle=0}}}
  \caption{Constraints on $w_{0}$ and $w_{1}$ from $1\sigma$ to $3\sigma$ with dark energy whose equation
  state is $w_{z}=w_{0}+w_{1}z/(1+z)^2$ in the top panel.
  The long dashed contours are derived from SNe Ia alone and the dotted contours
  are
derived from 20 GRBs and Gold sample. The diagonal lines are
obtained from 20 GRBs. The bottom panel shows the probability versus
transition redshift derived from the GRB and SN Ia sample.}
  \end{center}
  \end{figure}

\end{document}